\documentclass[twocolumn,pre,aps,nofootinbib]{revtex4}

\usepackage{graphicx}
\graphicspath{{./fig/}}
\usepackage{bm}

\begin{document}


%

\newcommand{\EQ}{\begin{equation}}
\newcommand{\EN}{\end{equation}}
\newcommand{\EQA}{\begin{eqnarray}}
\newcommand{\ENA}{\end{eqnarray}}
\newcommand{\eq}[1]{(\ref{#1})}
\newcommand{\EEq}[1]{Equation~(\ref{#1})}
\newcommand{\Eq}[1]{Eq.~(\ref{#1})}
\newcommand{\Eqs}[2]{Eqs~(\ref{#1}) and~(\ref{#2})}
\newcommand{\eqs}[2]{(\ref{#1}) and~(\ref{#2})}
\newcommand{\Eqss}[2]{Eqs~(\ref{#1})--(\ref{#2})}
\newcommand{\Sec}[1]{\S\,\ref{#1}}
\newcommand{\Secs}[2]{\S\S\,\ref{#1} and~\ref{#2}}
\newcommand{\Fig}[1]{Fig.~\ref{#1}}
\newcommand{\FFig}[1]{Figure~\ref{#1}}
\newcommand{\Tab}[1]{Table~\ref{#1}}
\newcommand{\Figs}[2]{Figs.~\ref{#1} and \ref{#2}}
\newcommand{\Tabs}[2]{Tables~\ref{#1} and \ref{#2}}
\newcommand{\bra}[1]{\langle #1\rangle}
\newcommand{\bbra}[1]{\left\langle #1\right\rangle}
\newcommand{\mean}[1]{\overline #1}
\newcommand{\meanB}{\overline{B}}
\newcommand{\mod}[1]{\mid\!\!#1\!\!\mid}
\newcommand{\chk}[1]{[{\em check: #1}]}

\newcommand{\Reynolds}{\mathrm{Re}}
\newcommand{\Rm}{\mathrm{Re}_{\rm M}}
\newcommand{\Pm}{\mathrm{Pr}_{\rm M}}
\newcommand{\ii}{\mathrm{i}}

%
%
\newcommand{\gggg}{\bm{g}}
\newcommand{\ddd}{\bm{d}}
\newcommand{\rrr}{\bm{r}}
\newcommand{\xx}{\bm{x}}
\newcommand{\yy}{\bm{y}}
\newcommand{\zzz}{\bm{z}}
\newcommand{\uu}{\bm{u}}
\newcommand{\vv}{\bm{v}}
\newcommand{\ww}{\bm{w}}
\newcommand{\mm}{\bm{m}}
\newcommand{\PP}{\bm{P}}
\newcommand{\QQ}{\bm{Q}}
\newcommand{\UU}{\bm{U}}
\newcommand{\bb}{\bm{b}}
\newcommand{\qq}{\bm{q}}
\newcommand{\BB}{\bm{B}}
\newcommand{\HH}{\bm{H}}
\newcommand{\II}{\bm{I}}
\newcommand{\AAA}{\bm{A}}
\newcommand{\aaa}{\bm{a}}
\newcommand{\aaaa}{\bm{a}} 
\newcommand{\eee}{\bm{e}}
\newcommand{\jj}{\bm{j}}
\newcommand{\JJ}{\bm{J}}
\newcommand{\nn}{\bm{n}}
\newcommand{\ee}{\bm{e}}
\newcommand{\ff}{\bm{f}}
\newcommand{\EE}{\bm{E}}
\newcommand{\FF}{\bm{F}}
\newcommand{\TT}{\bm{T}}
\newcommand{\CC}{\bm{C}}
\newcommand{\KK}{\bm{K}}
\newcommand{\MM}{\bm{M}}
\newcommand{\GG}{\bm{G}}
\newcommand{\kk}{\bm{k}}
\newcommand{\SSS}{\bm{S}}
\newcommand{\grav}{\bm{g}}
\newcommand{\nab}{\bm{\nabla}}
\newcommand{\OO}{\bm{\Omega}}
\newcommand{\oo}{\bm{\omega}}
\newcommand{\LL}{\bm{\Lambda}}
\newcommand{\llambda}{\bm{\lambda}}
\newcommand{\pomega}{\bm{\varpi}}
%
%
\newcommand{\RRRR}{\bm{\mathsf{R}}}
\newcommand{\SSSS}{\bm{\mathsf{S}}}

\newcommand{\DD}{\mathrm{D}}
\newcommand{\dd}{\mathrm{d}}

\newcommand{\onethird}{{\textstyle{\frac{1}{3}}}}
\newcommand{\W}{\,{\rm W}}
\newcommand{\V}{\,{\rm V}}
\newcommand{\kV}{\,{\rm kV}}
\newcommand{\T}{\,{\rm T}}
\newcommand{\G}{\,{\rm G}}
\newcommand{\Hz}{\,{\rm Hz}}
\newcommand{\kHz}{\,{\rm kHz}}
\newcommand{\kG}{\,{\rm kG}}
\newcommand{\K}{\,{\rm K}}
\newcommand{\g}{\,{\rm g}}
\newcommand{\s}{\,{\rm s}}
\newcommand{\ms}{\,{\rm ms}}
\newcommand{\cm}{\,{\rm cm}}
\newcommand{\m}{\,{\rm m}}
\newcommand{\km}{\,{\rm km}}
\newcommand{\kms}{\,{\rm km/s}}
\newcommand{\kg}{\,{\rm kg}}
\newcommand{\Mm}{\,{\rm Mm}}
\newcommand{\pc}{\,{\rm pc}}
\newcommand{\kpc}{\,{\rm kpc}}
\newcommand{\yr}{\,{\rm yr}}
\newcommand{\Myr}{\,{\rm Myr}}
\newcommand{\Gyr}{\,{\rm Gyr}}
\newcommand{\erg}{\,{\rm erg}}
\newcommand{\mol}{\,{\rm mol}}
\newcommand{\dyn}{\,{\rm dyn}}
\newcommand{\J}{\,{\rm J}}
\newcommand{\RM}{\,{\rm RM}}
\newcommand{\EM}{\,{\rm EM}}
\newcommand{\AU}{\,{\rm AU}}
\newcommand{\A}{\,{\rm A}}
%
%
\newcommand{\yan}[3]{, Astron. Nachr. {\bf #2}, #3 (#1).}
\newcommand{\yact}[3]{, Acta Astron. {\bf #2}, #3 (#1).}
\newcommand{\yana}[3]{, Astron. Astrophys. {\bf #2}, #3 (#1).}
\newcommand{\yanas}[3]{, Astron. Astrophys. Suppl. {\bf #2}, #3 (#1).}
\newcommand{\yanal}[3]{, Astron. Astrophys. Lett. {\bf #2}, #3 (#1).}
\newcommand{\yass}[3]{, Astrophys. Spa. Sci. {\bf #2}, #3 (#1).}
\newcommand{\ysci}[3]{, Science {\bf #2}, #3 (#1).}
\newcommand{\ysph}[3]{, Solar Phys. {\bf #2}, #3 (#1).}
\newcommand{\yjetp}[3]{, Sov. Phys. JETP {\bf #2}, #3 (#1).}
\newcommand{\yspd}[3]{, Sov. Phys. Dokl. {\bf #2}, #3 (#1).}
\newcommand{\ysov}[3]{, Sov. Astron. {\bf #2}, #3 (#1).}
\newcommand{\ysovl}[3]{, Sov. Astron. Letters {\bf #2}, #3 (#1).}
\newcommand{\ymn}[3]{, Monthly Notices Roy. Astron. Soc. {\bf #2}, #3 (#1).}
\newcommand{\yqjras}[3]{, Quart. J. Roy. Astron. Soc. {\bf #2}, #3 (#1).}
\newcommand{\ynat}[3]{, Nature {\bf #2}, #3 (#1).}
\newcommand{\sjfm}[2]{, J. Fluid Mech., submitted (#1).}
\newcommand{\pjfm}[2]{, J. Fluid Mech., in press (#1).}
\newcommand{\yjfm}[3]{, J. Fluid Mech. {\bf #2}, #3 (#1).}
\newcommand{\ypepi}[3]{, Phys. Earth Planet. Int. {\bf #2}, #3 (#1).}
\newcommand{\ypr}[3]{, Phys.\ Rev.\ {\bf #2}, #3 (#1).}
\newcommand{\yprl}[3]{, Phys.\ Rev.\ Lett.\ {\bf #2}, #3 (#1).}
\newcommand{\yepl}[3]{, Europhys. Lett. {\bf #2}, #3 (#1).}
\newcommand{\pcsf}[2]{, Chaos, Solitons \& Fractals, in press (#1).}
\newcommand{\ycsf}[3]{, Chaos, Solitons \& Fractals{\bf #2}, #3 (#1).}
\newcommand{\yprs}[3]{, Proc. Roy. Soc. Lond. {\bf #2}, #3 (#1).}
\newcommand{\yptrs}[3]{, Phil. Trans. Roy. Soc. {\bf #2}, #3 (#1).}
\newcommand{\yjcp}[3]{, J. Comp. Phys. {\bf #2}, #3 (#1).}
\newcommand{\yjgr}[3]{, J. Geophys. Res. {\bf #2}, #3 (#1).}
\newcommand{\ygrl}[3]{, Geophys. Res. Lett. {\bf #2}, #3 (#1).}
\newcommand{\yobs}[3]{, Observatory {\bf #2}, #3 (#1).}
\newcommand{\yaj}[3]{, Astronom. J. {\bf #2}, #3 (#1).}
\newcommand{\yapj}[3]{, Astrophys. J. {\bf #2}, #3 (#1).}
\newcommand{\yapjs}[3]{, Astrophys. J. Suppl. {\bf #2}, #3 (#1).}
\newcommand{\yapjl}[3]{, Astrophys. J. {\bf #2}, #3 (#1).}
\newcommand{\ypp}[3]{, Phys. Plasmas {\bf #2}, #3 (#1).}
\newcommand{\ypasj}[3]{, Publ. Astron. Soc. Japan {\bf #2}, #3 (#1).}
\newcommand{\ypac}[3]{, Publ. Astron. Soc. Pacific {\bf #2}, #3 (#1).}
\newcommand{\yannr}[3]{, Ann. Rev. Astron. Astrophys. {\bf #2}, #3 (#1).}
\newcommand{\yanar}[3]{, Astron. Astrophys. Rev. {\bf #2}, #3 (#1).}
\newcommand{\yanf}[3]{, Ann. Rev. Fluid Dyn. {\bf #2}, #3 (#1).}
\newcommand{\ypf}[3]{, Phys. Fluids {\bf #2}, #3 (#1).}
\newcommand{\yphy}[3]{, Physica {\bf #2}, #3 (#1).}
\newcommand{\ygafd}[3]{, Geophys. Astrophys. Fluid Dynam. {\bf #2}, #3 (#1).}
\newcommand{\yzfa}[3]{, Zeitschr. f. Astrophys. {\bf #2}, #3 (#1).}
\newcommand{\yptp}[3]{, Progr. Theor. Phys. {\bf #2}, #3 (#1).}
\newcommand{\yjour}[4]{, #2 {\bf #3}, #4 (#1).}
\newcommand{\pjour}[3]{, #2, in press (#1).}
\newcommand{\sjour}[3]{, #2, submitted (#1).}
\newcommand{\yprep}[2]{, #2, preprint (#1).}
\newcommand{\ppr}[2]{, Phys.\ Rev.\ {\bf #2}, in press (#1).}
\newcommand{\pproc}[3]{, (ed. #2), #3 (#1) (to appear).}
\newcommand{\yproc}[4]{, (ed. #3), pp. #2. #4 (#1).}
\newcommand{\ybook}[3]{, {\em #2}. #3 (#1).}

\title{Suppression of small scale dynamo action by an imposed magnetic field}

\author{Nils Erland L.\ Haugen}
  \affiliation{Department of Physics, The Norwegian University of Science
  and Technology, H{\o}yskoleringen 5, N-7034 Trondheim, Norway}
  \email{nils.haugen@phys.ntnu.no}
\author{Axel Brandenburg}
  \affiliation{NORDITA, Blegdamsvej 17, DK-2100 Copenhagen \O, Denmark}
  \email{brandenb@nordita.dk}

\date{\today,~ $ $Revision: 1.98 $ $}

\begin{abstract}
Non-helical hydromagnetic turbulence 
with an externally imposed magnetic field
is investigated using direct numerical simulations.
It is shown that
the imposed magnetic field lowers the spectral magnetic energy
in the inertial range.
This is explained by a suppression of the small scale dynamo.
At large scales, however,
the spectral magnetic energy increases with increasing imposed field
strength for moderately strong fields, and decreases only slightly for
even stronger fields.
The presence of Alfv\'en waves is explicitly confirmed by monitoring the
evolution of magnetic field and velocity at one point.
The frequency $\omega$ agrees with $v_{\rm A}k_1$, where $v_{\rm A}$ is the
Alfv\'en speed and $k_1$ is the smallest wavenumber in the box.
\end{abstract}
\maketitle

\section{Introduction}

Turbulent magnetic fields are seen in many astrophysical settings
\cite{Beck_etal96,BH98,Biskamp03}.
Such magnetic fields usually result from the conversion of kinetic energy
into magnetic energy, i.e.\ from dynamo action.
Numerical simulations show that a dynamo-generated magnetic field can be
of appreciable strength even when there is no kinetic helicity
\cite{CV00a,Scheko02}.
Simulations have recently also shown that at scales smaller than about five
times the energy carrying scale the magnetic energy spectrum seems to enter an
inertial subrange where the magnetic spectral energy exceeds the kinetic
spectral energy \cite{HBD03}.
This means that over any subvolume, whose scale is
within the inertial range, there is always a larger scale component
of the field with significant strength.
This raises the question whether one can model the small scale properties
of such turbulence simply by imposing a magnetic field.

A lot of work has already been devoted to studying hydromagnetic turbulence
in the presence of an external field \cite{CV00,MG01,CLV02}.
Nevertheless, the super-equipartition magnetic energy seen in simulations
without imposed field has never been seen in simulations with imposed field.
An exception is when the magnetic Prandtl number 
is large\cite{CLV02b}. However, the super-equipartition is then seen 
between the viscous and the resistive cutoff -- not in the inertial range.
It is one of our goals to elucidate this puzzle.
Likewise, although dynamos with helicity can produce substantial
super-equipartition on the scale of the system, they too are not able
to produce super-equipartition in the inertial range \cite{B01}.
In that sense the difference between dynamos with and without imposed field
is similar to the difference between helical and nonhelical dynamos.

The views on the effects of external fields are divided.
A common scenario that applies when the conditions for dynamo action
are not met (e.g.\ if the magnetic Reynolds number is too small), is one where
a local magnetic field can be enhanced simply by winding up an external
magnetic field.
Possible candidates where this may be the case are Io and Ganymede,
in which convection interacts with the field of Jupiter leading to
local field enhancement \cite{Schubert96,Khurana97,Sarson97}.
A similar possibility may also apply to the solar convection zone where
the large scale field of the 11 year solar cycle is primarily located at
the bottom of the convection zone \cite{SW80}, but the overlying convection
zone may shred the field to produce small scale field \cite{Sch84}.
Another possibility that has been discussed more recently is that the
small scale field at the solar surface could be generated locally
by a small scale dynamo operating near the surface \cite{Cat99}.

In hydromagnetic turbulence theory magnetic and kinetic energy are assumed
to cascade from large to small scales, similar to the hydrodynamic case,
although recent work has established a strong intrinsic anisotropy
\cite{GS95}, which has no counterpart in the hydrodynamic case.
However, this theory does not address the possibility of dynamo action.
It remains therefore an open question as to what is the nature of the interaction
resulting from imposed and dynamo-generated magnetic fields.
In particular, we shall present evidence that the imposed magnetic
field does not enhance dynamo action.
Instead, the external field
does actually suppress dynamo action, albeit in a subtle way
because the rms turbulent velocity is generally {\it not} decreased by
a modestly strong magnetic field.
We show that the suppression can be associated with the work term
resulting from the Lorentz force due to the imposed field.
It turns out that this term changes sign above a certain field strength
such that a certain fraction of magnetic energy flows backwards
to enhance the kinetic energy instead.

\section{Equations}

We adopt an
isothermal equation of state with constant (isothermal) sound speed $c_{\rm s}$,
so the pressure $p$ is related to the density $\rho$ by
$p=\rho c_{\rm s}^2$. The equation of motion is written in the form
\EQ
{\DD\uu\over\DD t}=-c_{\rm s}^2\nab\ln\rho+{\JJ\times\BB\over\rho}
+\FF_{\rm visc}+\ff,
\label{dudt}
\EN
where $\DD/\DD t=\partial/\partial t+\uu\cdot\nab$ is the advective
derivative, $\JJ=\nab\times\BB/\mu_0$ is the current density, $\mu_0$
is the vacuum permeability,
\EQ
\FF_{\rm visc}=\nu\left(\nabla^2\uu+\onethird\nab\nab\cdot\uu
+2\SSSS\cdot\nab\ln\rho\right),
\EN
is the viscous force,
$\nu=\text{const}$ is the kinematic viscosity,
\EQ
{\sf S}_{ij}=\frac{1}{2}\left({\partial u_i\over\partial x_j}
+ {\partial u_j\over\partial x_i}
-\frac{2}{3} \delta_{ij}\nab\cdot\uu\right)
\EN
is the traceless rate of strain tensor, and $\ff$ is a
random forcing function that consists of non-helical
plane waves; see Refs~\cite{HBD03,HBD04} for details.
The continuity equation is
written in terms of the logarithmic density,
\EQ
{\DD\ln\rho\over\DD t}=-\nab\cdot\uu,
\EN
and the induction equation is solved in terms of the magnetic vector
potential $\AAA$,
\EQ
{\partial\AAA\over\partial t}=\uu\times\BB+\eta\nabla^2\AAA,
\label{dAdt}
\EN
where $\eta=\text{const}$ is the magnetic diffusivity,
and $\BB=\BB_0+\bb$ is the magnetic field consisting
of the imposed uniform ($k=0$) field, $\BB_0$, and the deviations from the
imposed field $\bb=\nab\times\AAA$.
This split is necessary because the vector potential corresponding
to $\BB_0$ cannot be periodic, while both $\BB$ and $\AAA$
can well be assumed to be periodic.

\begin{table}[t!]
\caption{
Summary of the different runs with forcing at $k_{\rm f}=1.5$. 
All runs have magnetic Prandtl number unity.
The field strengths 0.06, 0.3, and 3.0 correspond roughly to
0.5, 2.0, and 20 times $B_{\rm eq}=\sqrt{\mu_0\rho_0}u_{\rm rms}$.
\label{Truns}}
\begin{ruledtabular}
\begin{tabular}{ccccccc}
Run&Res.     &$\nu=\eta$           &$\Rm$&$b_{\rm rms}$&    $u_{\rm rms}$ & $B_0$     \\
\hline
B4  &$128^3$  &$4   \times 10^{-4}$&     280   &      0.076 &0.17 &0.3 \\ 
C1  &$256^3$  &$2   \times 10^{-4}$&     400   &      0.062 &0.12 &0   \\ 
C2  &$256^3$  &$2   \times 10^{-4}$&     400   &      0.070 &0.12 &0.01\\ 
C3  &$256^3$  &$2   \times 10^{-4}$&     370   &      0.094 &0.12 &0.06\\ 
C4  &$256^3$  &$2   \times 10^{-4}$&     500   &      0.088 &0.19 &0.3 \\ 
C5  &$256^3$  &$2   \times 10^{-4}$&     500   &      0.075 &0.15 &3   \\ 
D4  &$512^3$  &$1   \times 10^{-4}$&     930   &      0.089 &0.14 &0.3 \\ 
E1  &$1024^3$ &$8   \times 10^{-5}$&    1000   &      0.075 &0.12 &0 \\ 
\end{tabular}
\end{ruledtabular}
\end{table}

In the simulations summarized in \Tab{Truns}
we have used the same method as described in
Ref.~\cite{HBD04}.
Kinetic and magnetic Reynolds numbers are defined as
\EQ
\text{Re}={u_{\rm rms}\over\nu k_{\rm f}},\quad
\Rm={u_{\rm rms}\over\eta k_{\rm f}},
\EN
respectively.
Here, $k_{\rm f}$ is the average forcing wavenumber and
$\Pm=\nu/\eta\equiv\Rm/\text{Re}$ is the magnetic Prandtl number.
In all cases studied below we assume $\Pm=1$.
We study cases where $k_{\rm f}$ is either 1.5 or 5.

The Pencil Code \cite{PC} is used for all our simulations.
The resolution is varied between $128^3$ and $1024^3$ meshpoints.
Although we solve the compressible equations, the sound speed
is large compared with the turbulent velocities.
We find that the energies of
solenoidal and potential components of the flow have a ratio
$E_{\rm pot}/E_{\rm sol} \approx 10^{-4}\mbox{--}10^{-2}$ for most scales;
only towards the Nyquist frequency the ratio increases to about $0.1$.
Thus, our results should be close to the incompressible limit.

We use non-dimensional quantities by measuring length in units of
$1/k_1$ (where $k_1=2\pi/L$ is the smallest wavenumber in the box of size
$L$; in the present case $L=2\pi$),
speed in units of the isothermal sound speed $c_{\rm s}$, density
in units of the initial value $\rho_0$, and magnetic field in
units of $(\mu_0\rho_0 c_{\rm s}^2)^{1/2}$.

\section{Energy balance}

\begin{figure}[t!]\centering\includegraphics[width=0.40\textwidth]
{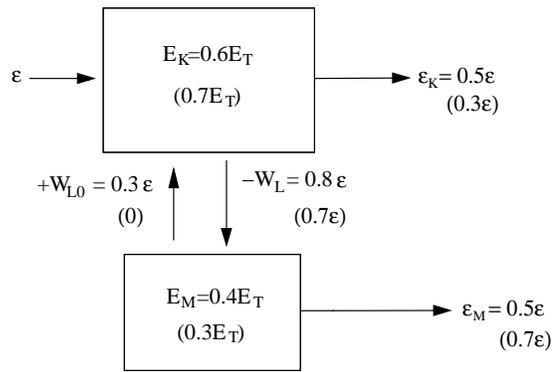}\caption{
Sketch of the energy budget
showing the kinetic and magnetic energy reservoirs
together with the flow of energy for Run~D4. The numbers in 
parentheses correspond to Run~E1 without an imposed field.
The total dissipation rate is denoted by
$\epsilon\equiv\epsilon_{\rm K}+\epsilon_{\rm M}$, which is
the sum of kinetic and magnetic energy dissipation rates.
The $+$~sign on $W_{\rm L0}$ and the direction of the corresponding
arrow emphasize that, at least for sufficiently strong imposed fields, energy
flows from the magnetic to the kinetic energy reservoir.
}\label{energy_sceme}\end{figure}

A sketch of the overall energy budget is given in \Fig{energy_sceme}
where we show the magnetic and kinetic energy reservoirs
together with arrows indicating the flow of energy.
The arrow pointing
into the kinetic energy reservoir is the energy flux $\epsilon$
entering the simulation through the external forcing, while the 
arrows pointing to the right from the kinetic and magnetic energy
reservoirs denote viscous and Joule dissipation,
i.e.\ $\epsilon_{\rm K}$ and $\epsilon_{\rm M}$, respectively.
On the average and in the statistically steady state
we expect $\epsilon=\epsilon_{\rm K}+\epsilon_{\rm M}$.

The two arrows between the kinetic and magnetic energy
reservoirs correspond to the contributions to the work done against
the Lorentz force.
In general this work term can be written as $\bra{\uu\cdot(\jj\times\BB)}$,
where we have used $\jj=\JJ$ to emphasize that the current density has
vanishing volume average.
However, since $\BB=\BB_0+\bb$, where $\bb=\nab\times\AAA$ is the departure
from the imposed field (also with vanishing volume average),
we can divide the work term into a contribution from the
fluctuating field, $\bra{\uu\cdot(\jj\times\bb)}$, and one
from the imposed field, $\bra{\uu\cdot(\jj\times\BB_0)}$.
The latter can also be written as $\BB_0\cdot\bra{\uu\times\jj}$,
which emphasizes the fact that this term is quadratic in the
fluctuations and can hence only transfer energy between kinetic and magnetic
energy reservoirs at the same wavenumber.
We can thus write
\begin{eqnarray}
\frac{\dd E_{\rm M}}{\dd t}=-W_{\rm L}-W_{\rm L0}-\epsilon_{\rm M},
\label{dEmag}
\end{eqnarray}
where $E_{\rm M}=\bra{\bb^2}/\mu_0$ is the magnetic energy
(per unit volume) of the induced field without the imposed field,
$W_{\rm L}=\langle\uu\cdot(\jj\times\bb)\rangle$ is the work done
by the fluctuating fields,
$W_{\rm L0}=\BB_0\cdot\langle\uu\times\jj\rangle$ is the work done
against winding up the imposed mean field, and
$\epsilon_{\rm M}=\eta\mu_0\bra{\jj^2}$ is the loss from Joule heating.
In a closed or periodic system such as the one considered here,
there are no surface terms, which is why there is no term associated
with the Poynting flux in \Eq{dEmag}.

\begin{figure}[t!]\centering\includegraphics[width=0.40\textwidth]
{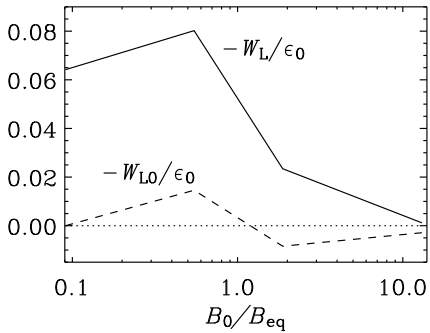}\caption{
Dependence of $-W_{\rm L}$ (solid line) and
$-W_{\rm L0}$ (dashed line) on the imposed
field strength, $B_0$. Here $B_{\rm eq}=\sqrt{\mu_0\rho_0}u_{\rm rms}$ is
the equipartition field strength and $\epsilon_0=k_{\rm f}\rho_0 u_{\rm rms}^3$
is a reference value for the energy flux. $k_{\rm f}=1.5$.
}\label{suppression2}\end{figure}

The numbers on the arrows give the energy fluxes
for a simulation with a moderately strong imposed field (Run~D4). 
The numbers in parentheses are the corresponding values
for a simulation without imposed fields.
By comparing the two we see that with an imposed field the content of
the magnetic energy reservoir is slightly increased.
Nevertheless,
magnetic dissipation has decreased and kinetic dissipation has increased.
This suggests that an imposed magnetic field quenches the dynamo.

Naively one might have expected that the $-W_{\rm L0}$ term always ``helps''
the dynamo and that it therefore always transfers energy from kinetic to
magnetic energy by winding up the imposed field.
This is however not the case.
In \Fig{suppression2} we show $W_{\rm L}$ and $W_{\rm L0}$,
normalized by $\epsilon_0=k_{\rm f}\rho_0 u_{\rm rms}^3$, as functions 
of imposed field strength.
(In those units the total energy input to the system is
$\epsilon\approx0.07\epsilon_0$.)
The negative contribution from 
$-W_{\rm L0}$ for large field strengths is actually
the main reason that simulations
with strong imposed fields have less magnetic 
energy; see Table \ref{Truns}.
Since the $W_{\rm L0}$ term is local in $k$-space 
it also explains the general increase in kinetic energy at all scales.

\begin{figure}[t!]\centering\includegraphics[width=0.40\textwidth]
{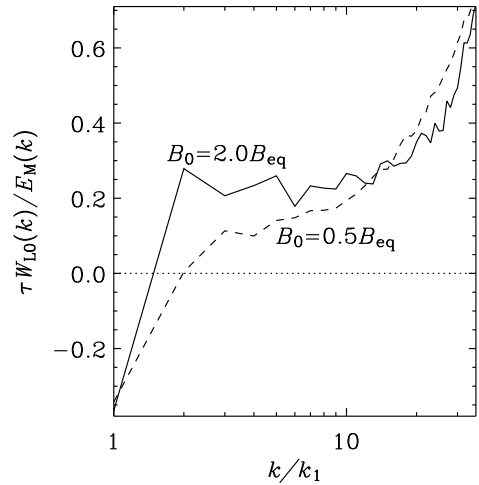}\caption{
Contribution to the
spectral energy transfer between kinetic and magnetic energies
due to the imposed field.
Power spectrum of $\BB_0\cdot\bra{\uu_k\times\jj_k}$ normalized by
$E^M_k$ for run~C4 (solid line) and run~C3 (dashed line). 
We clearly see that at the box scale there is transport of
energy from kinetic to magnetic field, while at all other scales the 
transport is in the opposite direction, i.e. there is a suppression of 
the magnetic field. It is also clear that the suppression is much 
stronger at the smallest scales.
}\label{uxj}\end{figure}

To quantify the above statement, we discuss now the spectral
energy transfer function, $W_{\rm L0}(k)=\BB_0\cdot\bra{\uu_k\times\jj_k}$,
where $\uu_k$ and $\jj_k$ are Fourier filtered velocity and current density.
In \Fig{uxj} we plot the ratio of $W_{\rm L0}(k)$ to the magnetic energy
spectrum $E_{\rm M}(k)$, divided by the eddy turnover time $\tau=(k_{\rm
f}u_{\rm rms})^{-1}$, using data from Runs~C3 and C4.
Here, the spectra are normalized such that
$\int W_{\rm L0}(k)\,\dd k=W_{\rm L0}$ and
$\int E_{\rm M}(k)\,\dd k=E_{\rm M}$.
It turns out that, first, $W_{\rm L0}(k)$ has a positive contribution to
the magnetic energy at small wavenumbers.
This explains the increase in magnetic energy at large scales
(small wavenumbers).
Second, at moderate and large wavenumbers, $W_{\rm L0}(k)$ is positive,
which explains the suppression of the magnetic energy.

\section{Spectral energy changes}

\begin{figure}[t!]\centering\includegraphics[width=0.40\textwidth]
{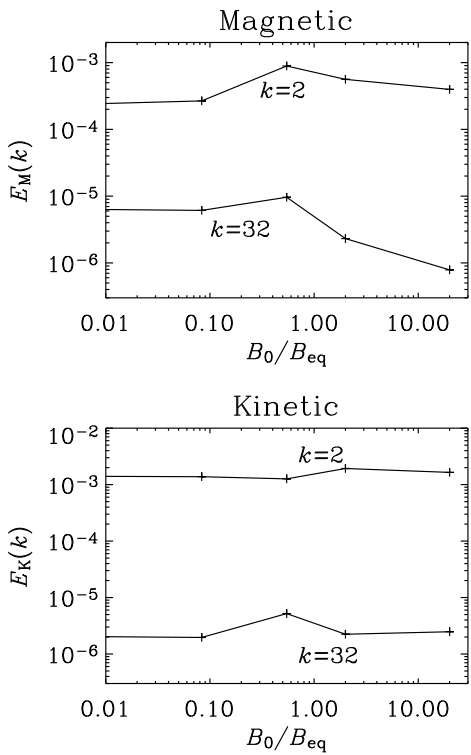}\caption{
Magnetic (top) and kinetic (bottom) spectral energy at wavenumbers
2 and 32 as a function of $B_0$. $k_{\rm f}=1.5$.
}\label{E_B0}\end{figure}

\begin{figure}[t!]\centering\includegraphics[width=0.40\textwidth]
{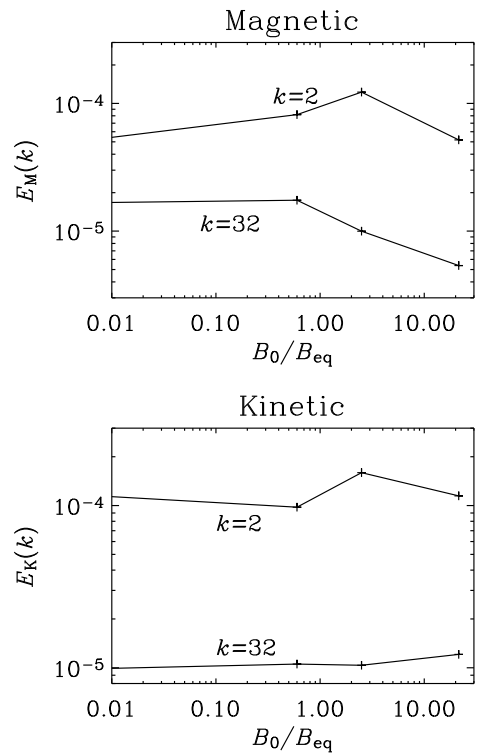}\caption{
Same as \Fig{E_B0} but with forcing at $k_{\rm f}=5$.
}\label{E_B0_kf5}\end{figure}

Next we investigate
the effect of varying the strength of the imposed field on the
magnetic and kinetic energies at different
wavenumbers; see \Fig{E_B0}.
We see that around $B_0\approx B_{\rm eq}$ the magnetic energy is somewhat
enhanced at small wavenumbers ($k=2$, corresponding to modestly large scales),
but decreased at large wavenumbers ($k=32$, corresponding to small scales).
At the same time the effect on the velocity field is weak, but there
is generally a tendency for enhanced velocities, especially at large scales.

When the forcing is at $k_{\rm f}=5$, instead of at $k_{\rm f}=1.5$,
the trends are very similar as in \Fig{E_B0}; see \Fig{E_B0_kf5}.
In particular, at large scales there is first an increase and then a decrease 
of the magnetic energy as the imposed field strength is increased, while
for small scales the magnetic energy decreases for all imposed field
strengths.
We therefore conclude that the suppression of the magnetic field is, 
at least qualitatively, independent of the forcing scale.

The situation is different in the presence of helicity where
it has been argued that it
is particularly the large scale magnetic field at $k=k_1$
that is affected by the
introduction of an imposed magnetic field \cite{MMMD02}.
This can be interpreted as a suppression of the $\alpha$ effect.
Repeating the simulations of Ref.~\cite{MMMD02},
we were able to confirm their findings; see also Ref.~\cite{BM04}.
We also find that kinetic and magnetic energy spectra fall almost
on top of each other when $B_0=2B_{\rm eq}$.
This is just like in the case without imposed field \cite{B01}, except at
$k=k_1$ where there is an additional field component due to the $\alpha$
effect.
Furthermore, increasing the field to $B_0=20\,B_{\rm eq}$ we do recover 
the same suppression of the dynamo as without helicity,
i.e.\ the spectra look similar to those of Run~C5.
Thus, the suppression of dynamo activity by the imposed field is
rather general and affects equally helical and nonhelical dynamos.

\begin{figure}[t!]\centering\includegraphics[width=0.40\textwidth]
{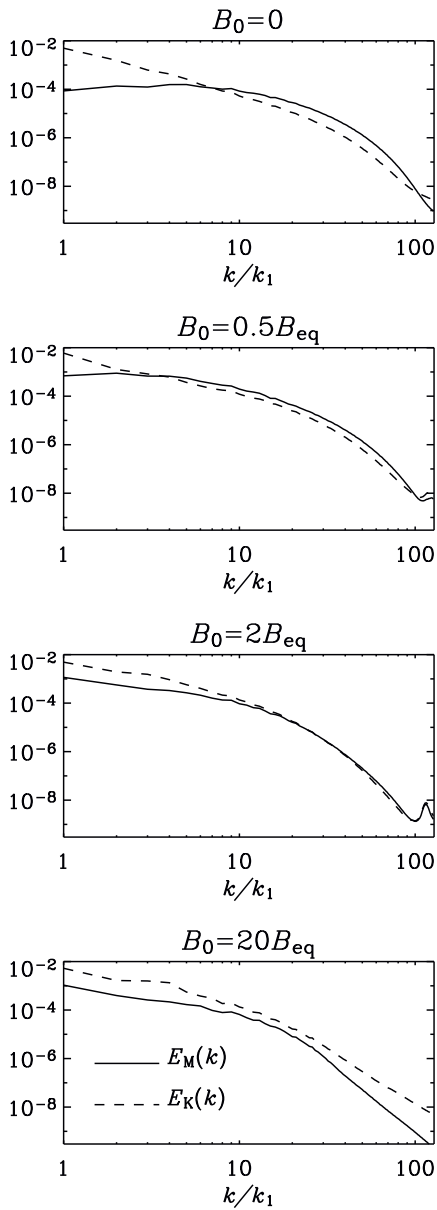}\caption{
Magnetic and kinetic energy spectra for
runs with different imposed field strengths (Runs C1 and C3-C5).
In all cases $B_{\rm eq}=0.12$--$0.15$; see \Tab{Truns}.
}\label{power_comp_externalB_256}\end{figure}

It is generally believed that hydromagnetic turbulence can be
described as an ensemble of Alfv\'en waves.
This is true both for the Goldreich-Sridhar \cite{GS95} 
and Iroshnikov-Kraichnan \cite{Iro63,Kra65} theories.
This would then suggest that magnetic and kinetic energies should be
comparable to one another at each scale. From
\Figs{power_comp_externalB_256}{power_comp_externalB_256_kf5} 
we see that magnetic and kinetic energies are close to each other,
but generally not equal.
This is also seen in the simulations of Cho \& Vishniac \cite{CV00}.
Only when the imposed field is approximately equal in strength 
to the rms field do we have approximate equipartition between magnetic
and kinetic energies at small scales.

\section{Shape of the energy spectra}

\begin{figure}[t!]\centering\includegraphics[width=0.40\textwidth]
{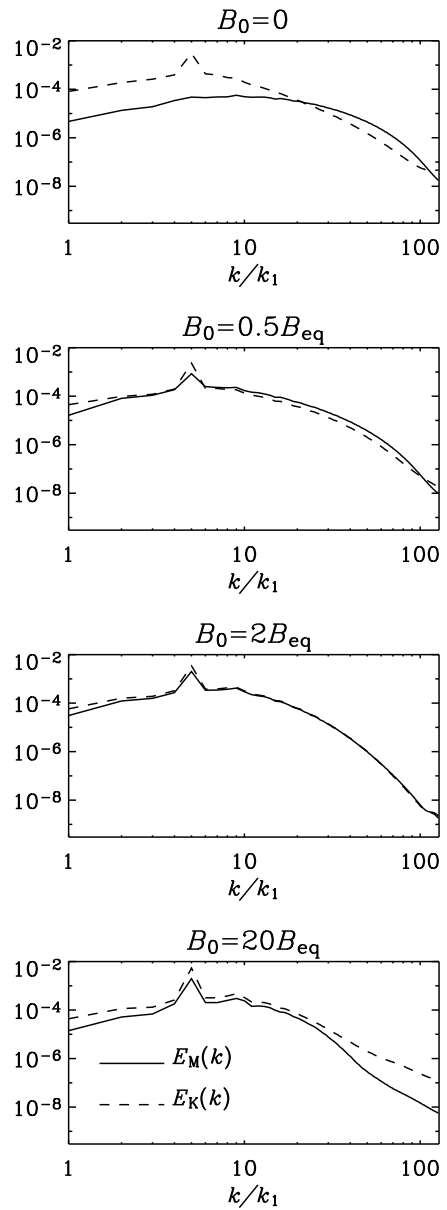}\caption{
Magnetic and kinetic energy spectra for 
runs with different imposed field strengths and forcing at $k=5$.
}\label{power_comp_externalB_256_kf5}\end{figure}

\begin{figure}[t!]\centering\includegraphics[width=0.45\textwidth]
{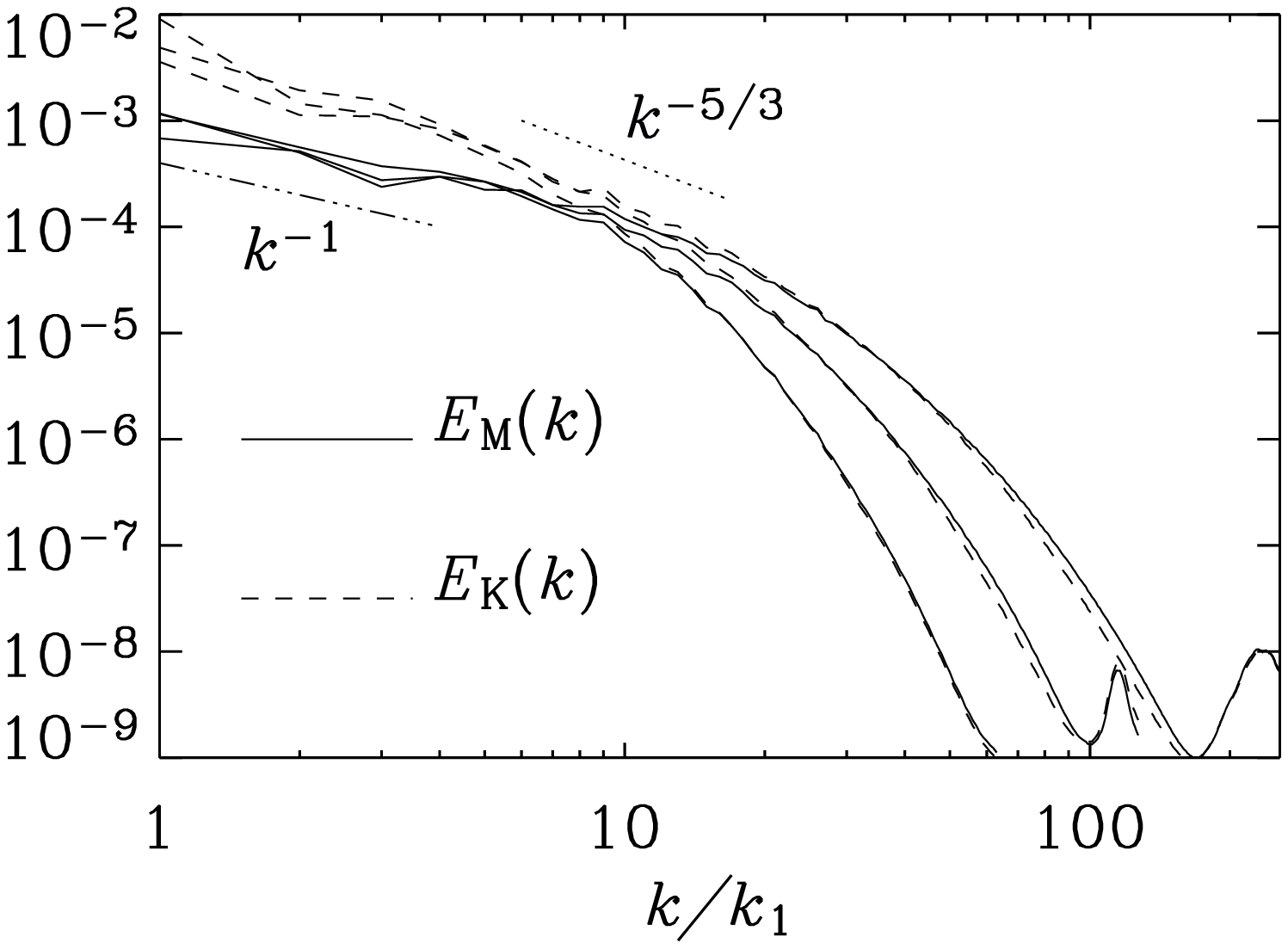}\caption{
Magnetic and kinetic power spectra for runs with $B_0=0.3$
(runs B4, C4 and D4).
}\label{B=0.3_diffRe}\end{figure}

As the resolution is increased, one begins to see indications of the
build-up of a short $k^{-5/3}$ inertial range of kinetic and magnetic
energies at intermediate wavenumbers; see \Fig{B=0.3_diffRe}.
The inertial range is as yet too short to be
conclusive, and we therefore need larger simulations in
order to be sure whether we have a real $k^{-5/3}$ slope or not.

From \Fig{B=0.3_diffRe} we also see that in the range $k_1<k<10$ the 
magnetic energy spectrum seems to follow a $k^{-1}$ slope.
For comparison, in the case without an imposed field the spectral magnetic
energy was actually increasing with $k$ and followed approximately a
$k^{1/3}$ slope \cite{HBD04} at small $k$.
The $k^{-1}$ spectrum for imposed fields can be motivated by 
dimensional arguments:
assume that the magnetic energy spectrum is a function of the imposed
field strength $B_0$ and the wavenumber $k$, and that the spectrum
is given by the ansatz $E_{\rm M}(k)=C B_0^a k^b$,
then, from dimensional arguments, one finds $a=2$ and $b=-1$, so
\EQ
\label{k-1}
E_{\rm M}(k)=C B_0^2 k^{-1}
\label{kminus1spectrum}
\EN
where $C$ is a dimensionless constant.
Such a spectrum is expected if there is a mean field \cite{KR94},
but it may generally also appear at the low wavenumber end of the
inertial subrange \cite{RS82}, and indications of this spectrum
have been seen in convective dynamo simulations \cite{BJNRST96}.
It turns out, however, that the value of $C$ (obtained from a fit)
is different for different values of $B_0$, casting doubt on the
validity of the assumptions behind \Eq{kminus1spectrum}.
We therefore 
discard this simple explanation of the large scale magnetic spectrum.
Indeed in \Fig{power_comp_externalB_256_kf5}  we
see that we get no longer a $k^{-1}$ magnetic energy spectrum for 
large scales when the forcing is at $k=5$; 
instead the infrared part of the spectrum has an increasing slope 
close to $E_{\rm M}(k)\sim k$ for $k<k_{\rm f}$.
Some intermediate behavior is seen when $k_{\rm f}=2...3$;
see Ref.~\cite{CV00} where no $k^{-1}$ behavior was found.

\section{Direct evidence for Alfv\'en waves}

\begin{figure}[t!]\centering\includegraphics[width=0.45\textwidth]
{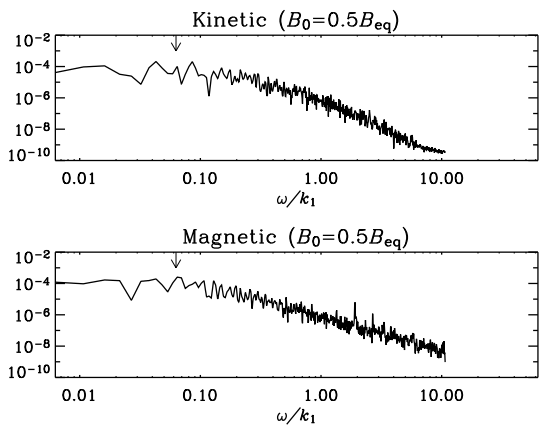}\caption{Fourier spectra of time evolution of 
the magnetic and kinetic fields at one point in the box for
simulation with $B_0=0.06$.
The point of interest is chosen to be in the center of the box.
The arrows represent the frequency of an Alfv\'en wave with a wavelength of
the box size traveling along the imposed field. We clearly see that 
Alfv\'en waves are strongly present in the simulation.
}\label{power_time_evol0.06}\end{figure}

\begin{figure}[t!]\centering\includegraphics[width=0.45\textwidth]
{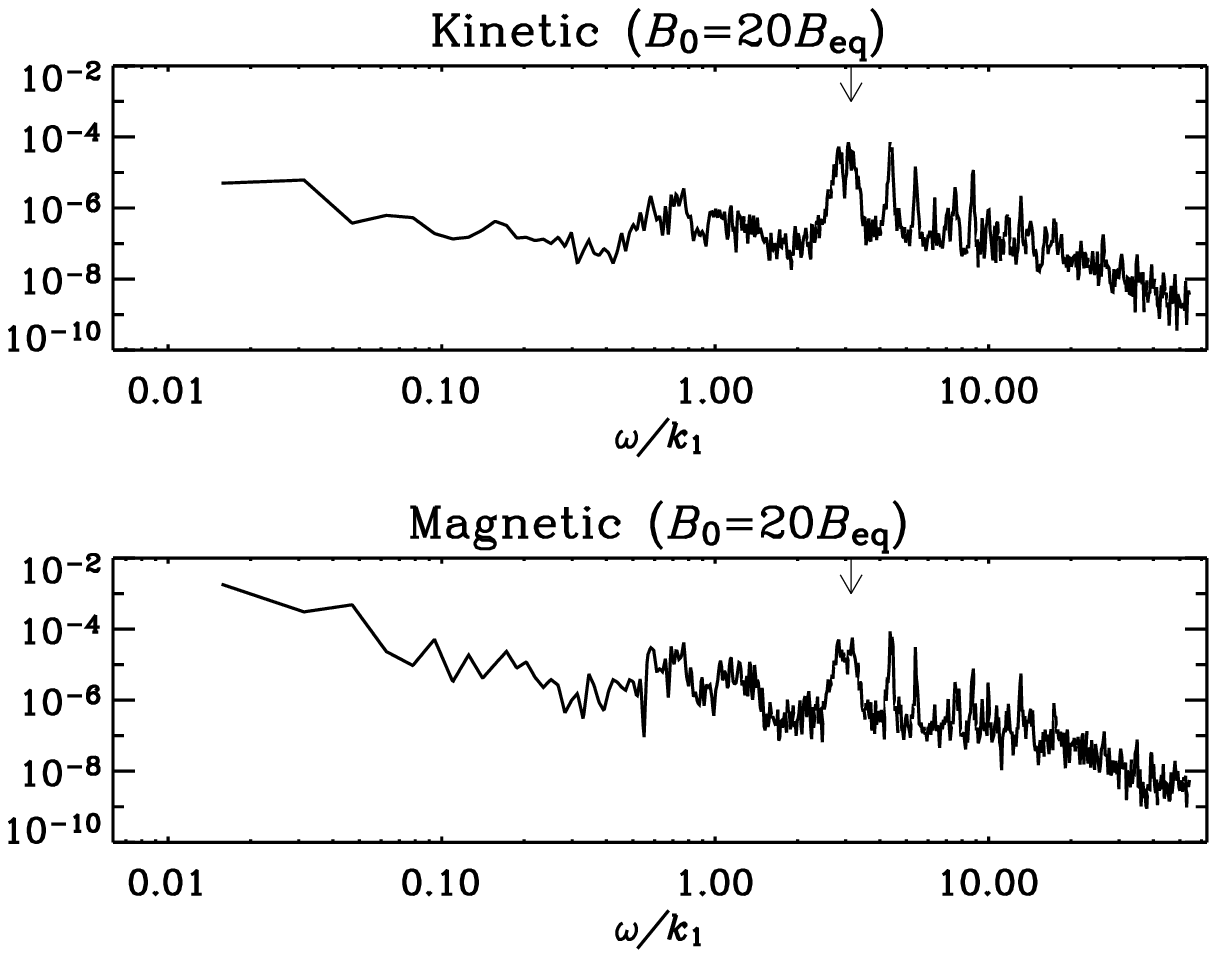}\caption{
Same as \Fig{power_time_evol0.06} but with $B_0=3.0$.
}\label{power_time_evol3.0}\end{figure}

Finally we look at the frequency power spectrum calculated from the time series
of the magnetic field and velocity at one point in the simulation box;
see \Figs{power_time_evol0.06}{power_time_evol3.0}.
As expected, the larger
the imposed magnetic field, the faster does the field oscillate.
The peaks in the power spectra for $B_0=3.0$ and $B_0=0.06$ correspond
to the frequency of the corresponding Alfv\'en wave,
\EQ
\omega=v_{\rm A}k_1,
\quad\mbox{where}\quad v_{\rm A}=B_0/\sqrt{\mu_0\rho_0}
\EN 
is the Alfv\'en speed.
(In our case we have $\mu_0=\rho_0=1$.)
When $B_0$ is comparable to or less than $B_{\rm eq}$, the peaks in the spectra
are no longer well pronounced.

For strong fields, however, the Alfv\'en peaks are seen quite clearly.
It is conceivable that these Alfv\'en waves are stochastically excited
by the turbulence.
This might be similar to the stochastic driving of acoustic waves in the
solar convection zone \cite{GP90}.

\section{Conclusion}

The present studies have shown that a uniformly imposed magnetic field
has two important effects on the magnetic field that is induced at
finite wavenumbers ($k\neq0$).
First, the magnetic field is slightly enhanced at and around the forcing
wavenumber (corresponding to the energy carrying scale).
Second, the magnetic field is quenched with increasing $B_0$ at all
larger wavenumbers corresponding to the inertial and diffusive subranges.

The enhancement and suppression at the two different wavenumber ranges is
associated with a corresponding wavenumber dependence of the work term,
$\BB_0\cdot\bra{\uu\times\jj}$.
The suppression of the magnetic field in the inertial range is quite
opposite to the behavior without imposed field when there is instead a
significant enhancement of the magnetic energy spectrum over the kinetic
energy spectrum.
We therefore refer to this effect as a suppression of the dynamo by the
imposed field.

The suppression of dynamo activity
might be a consequence of the tendency toward two-dimensionalization of
the turbulence by the large scale field \cite{MBG03}.
Such an effect is well-known for low-$\Rm$ hydromagnetic turbulence
\cite{Schumann76}, and it is a mathematical theorem that there can be no
dynamo action in two dimensions \cite{Zeldovich57}.
Of course, the turbulence does not really become two-dimensional, but
instead the correlation length along the field becomes large.
This type of anisotropy is a crucial ingredient of the
Goldreich-Sridhar theory \cite{GS95}.

The Goldreich-Sridhar
theory also predicts that Alfven waves should be present in the system.
This has been confirmed by inspecting velocities and
magnetic fields at a single point in the middle of the simulation box.
These Alfv\'en waves have the expected frequency $\omega_A=v_{\rm A}k_1$.
Furthermore, we do not find that there is equipartition 
between magnetic and kinetic energy spectra in the inertial range for large 
imposed field strengths. 
The absence of equipartition may be a consequence of the inertial range
being still too short (or absent).
In runs where $B_0=B_{\rm eq}$, on the other hand, there is clear evidence
that kinetic and magnetic energy spectra fall on top of each other
throughout the dissipation subrange.
This is also in agreement with earlier results of Cho and collaborators
\cite{CLV02}, who considered the case where the imposed field had
equipartition strength.

Whether or not models with imposed field can reproduce the situation
in small sub-domains of simulations with no overall imposed field is
still unclear.
At first glance the answer seems to be no, because none of the
simulations with imposed field have ever been able to produce
super-equipartition in the inertial range, as it is seen in the
nonhelical simulations without imposed field \cite{HBD03}.
However, the reason for this may well lie in the still insufficient
resolution of the simulations with no imposed field -- even though they
do already have a resolution of $1024^3$ meshpoints.
It is indeed possible that, even though the kinetic and magnetic energy
spectra are approximately parallel to each other over a certain range
of wavenumbers and offset by a factor of about 2.5, they may actually
converge at still larger wavenumbers.
Preliminary indications of this have now been seen in simulations using
hyperviscosity and hyper-resistivity with no imposed field.
However, a general difficulty with hyperviscosity and hyper-resistivity
is that certain aspects of the physics of such systems are significantly
modified \cite{BS02}.
It is therefore equally important to assess the features that are likely
not to be altered by this manipulation.
A detailed discussion of this will be the subject of a forthcoming paper.

\acknowledgments
We thank Tarek Yousef and Eric Blackman for valuable discussions and
comments on the manuscript, the Danish Center
for Scientific Computing for granting time on the Horseshoe cluster,
and the Norwegian High Performance Computing Consortium (NOTUR)
for granting time on the parallel computers in 
Trondheim (Gridur/Embla) and Bergen (Fire).



\begin{thebibliography}{}

\bibitem{Beck_etal96}
R. Beck, A. Brandenburg, D. Moss, A. Shukurov,
and D. Sokoloff\yannr{1996}{34}{155}

\bibitem{BH98}
S. A. Balbus and J. F. Hawley\yjour{1998}{Rev. Mod. Phys.}{70}{1}

\bibitem{Biskamp03}
D. Biskamp\ybook{2003}{Magnetohydrodynamic Turbulence}
{Cambridge: Cambridge University Press}

\bibitem{CV00a}
J. Cho and E. Vishniac\yapj{2000}{538}{217}

\bibitem{Scheko02}
A. Schekochihin, S. Cowley, G. W. Hammett, J. L. Maron
\& J. C. McWilliams\yjour{2002}{New J.\ Physics}{4}{84.1}

\bibitem{HBD03}
N. E. L. Haugen, A. Brandenburg, and W. Dobler\yapjl{2003}{597}{L141}

\bibitem{CV00}
J. Cho and E. T. Vishniac\yapj{2000}{539}{273}

\bibitem{MG01}
J. Maron and P. Goldreich\yapj{2001}{554}{1175}

\bibitem{CLV02}
Cho, J., Lazarian, A., \& Vishniac, E.\yapj{2002}{564}{291}

\bibitem{CLV02b}
Cho, J., Lazarian, A., \& Vishniac, E.\yapj{2002}{566}{L49}

\bibitem{B01}
A. Brandenburg\yapj{2001}{550}{824}

\bibitem{Schubert96}
G. Schubert, K. Zhang, M. G. Kivelson, J. D. Anderson\ynat{1996}{384}{544}

\bibitem{Khurana97}
K. K. Khurana, M. G. Kivelson, C. T. Russell\ygrl{1997}{24}{2391}

\bibitem{Sarson97}
G. R. Sarson, C. A. Jones, K. Zhang, G. Schubert\ysci{1997}{276}{1106}

\bibitem{SW80}
E. A. Spiegel and N. O. Weiss\ynat{1980}{287}{616}

\bibitem{Sch84}
M. Sch\"ussler\yproc{1984}{67}
{The Hydromagnetics of the Sun}
{T. D. Guyenne \& J. J. Hunt}{ESA SP-220}

\bibitem{Cat99}
F. Cattaneo\yapj{1999}{515}{L39}

\bibitem{GS95}
P. Goldreich and S. Sridhar\yapj{1995}{438}{763}

\bibitem{HBD04}
N. E. L. Haugen, A. Brandenburg, and W. Dobler\ypr{2004}{E 70}{016308}

\bibitem{PC}
The Pencil Code is a grid based high order code (sixth
order in space and third order in time) for solving the compressible
MHD equations; \url{http://www.nordita.dk/data/brandenb/pencil-code}.

\bibitem{MMMD02}
D. C. Montgomery, W. H. Matthaeus, L. J. Milano, and
P. Dmitruk\ypp{2002}{9}{1221}

\bibitem{BM04}
A. Brandenburg and W. H. Matthaeus,
Phys. Rev. {\bf E}, in press (2004);
see also {\sf astro-ph/0305373}.

\bibitem{Iro63}
R. S. Iroshnikov\ysov{1963}{7}{566}

\bibitem{Kra65}
R. H. Kraichnan\ypf{1965}{8}{1385}

\bibitem{KR94}
N. Kleeorin and I. Rogachevskii\ypr{1994}{E 50}{2716}

\bibitem{RS82}
A. A. Ruzmaikin and A. M. Shukurov\yass{1982}{82}{397}

\bibitem{BJNRST96}
A. Brandenburg, R. L. Jennings, \AA. Nordlund,
M. Rieutord, R. F. Stein, and I. Tuominen\yjfm{1996}{306}{325}

\bibitem{GP90}
P. Goldreich and P. Kumar\yapj{1990}{363}{694}

\bibitem{MBG03}
W.C. Muller, D. Biskamp, and R. Grappin\ypr{2003}{E 67}{066302}

\bibitem{Schumann76}
U. Schumann\yjfm{1976}{74}{31}

\bibitem{Zeldovich57}
Ya. B. Zeldovich\yjetp{1957}{4}{460}

\bibitem{BS02}
A. Brandenburg and G. S. Sarson\yprl{2002}{88}{055003}

\end{thebibliography}
\end{document}